\documentclass{llncs}
\usepackage{graphicx}
\usepackage{color}
\usepackage{url}
\usepackage{bchart}
\usepackage{array}
\usepackage{wasysym}

\newcolumntype{R}[1]{>{\raggedleft\arraybackslash}p{#1}}

\begin{document}

\title{
  Broadening the Scope of Nanopublications\thanks{
    The work presented in this paper has been supported by the National Library of Medicine grant 5R01LM009956.}
}

\author{
  Tobias Kuhn,\inst{1,2} Paolo Emilio Barbano,\inst{3} Mate Levente Nagy,\inst{4}\\Michael Krauthammer\inst{4,1}
}

\institute{
  Department of Pathology, Yale University School of Medicine\\
\and
  Chair of Sociology, in particular of Modeling and Simulation, ETH Zurich\\
\and
  Department of Mathematics, Yale University\\
\and
  Program for Computational Biology and Bioinformatics, Yale University
\smallskip\\
  \texttt{kuhntobias@gmail.com},
  \texttt{mate.nagy@yale.edu},
  \texttt{paoloemilio.barbano@yale.edu},
  \texttt{michael.krauthammer@yale.edu}
}

\maketitle

\begin{abstract}
In this paper, we present an approach for extending the existing concept of nanopublications --- tiny entities of scientific results in RDF representation --- to broaden their application range. The proposed extension uses English sentences to represent informal and underspecified scientific claims. These sentences follow a syntactic and semantic scheme that we call AIDA (Atomic, Independent, Declarative, Absolute), which provides a uniform and succinct representation of scientific assertions.
Such AIDA nanopublications are compatible with the existing nanopublication concept and enjoy most of its advantages such as information sharing, interlinking of scientific findings, and detailed attribution, while being more flexible and applicable to a much wider range of scientific results.
We show that users are able to create AIDA sentences for given scientific results quickly and at high quality, and that it is feasible to automatically extract and interlink AIDA nanopublications from existing unstructured data sources.
To demonstrate our approach, a web-based interface is introduced, which also exemplifies the use of nanopublications for non-scientific content, including meta-nanopublications that describe other nanopublications.
\end{abstract}

\section{Introduction}

Nanopublications have been proposed to make it easier to find, connect and curate core scientific statements and to determine their attribution, quality and provenance \cite{groth2010isu}. Small RDF-based data snippets --- i.e. nanopublications --- rather than classical narrative articles should be at the center of general scholarly communication \cite{mons2011naturegen}. In contrast to narrative articles, nanopublications support data sharing and mining, allow for fine-grained citation metrics on the level of individual claims, and give incentives for crowdsourced community efforts. In this paper, we propose an extension that allows for informal and underspecified representations and broadens the scope of the nanopublication approach.

The novelty of nanopublications lies in the combination of four ideas: (1) to subdivide scientific results into minimal pieces, (2) to represent these results --- called \emph{assertions} --- in an RDF-based formal notation, (3) to attach RDF-based provenance information on this ``atomic'' level, and (4) to treat each of these tiny entities as a separate publication.
Number (2) strikes us as problematic: Requiring formal representations for scientific results seems to be unrealistic in many cases and might restrict the range of practical application considerably. On the other hand, we think that the approach would be highly beneficial even if this restriction is dropped, and that at the same time it would become much more broadly applicable.
Specifically, we propose to allow authors to attach English sentences to nanopublications, thus allowing for informal representations of scientific claims. We previously sketched this approach in a position paper \cite{kuhn2012wole}.

To illustrate and motivate our approach, let us consider the following fictitious scenario: Giuseppe is a researcher in the biomedical area. Just now, he came to think of the possibility that gene X might accelerate the late stage of the course of disease Y, and he decides to investigate this further. The first questions that Giuseppe faces are: Has somebody else thought of gene X as a late-stage accelerator of disease Y? If so, is it an established fact or an open, maybe even controversial question? How much evidence is there on either side (i.e. the statement being true vs. false)? Who has worked on this question and what are their positions? With the current Web, it takes Giuseppe hours, probably days to answer these questions.
As it so happens, a researcher named Isabelle is asking herself the same question. One of her experiments, designed for an entirely different purpose, showed some evidence that gene X might speed up the final stage of disease Y. She wonders whether this would be a new finding or not, but she only has time for a quick Web search, which does not reveal anything.

With our approach, this scenario would turn out differently in the future. Giuseppe and Isabelle would each access a nanopublication portal to enter their hypothesis ``gene X accelerates the late stage of the course of disease Y.'' The system would retrieve related nanopublications, in particular those with matching sentences, including the ones that use different wording to express the same meaning (applying a mixture of automatic clustering and crowdsourcing).
In an instant, the system would compile and present the relevant information: the amount of existing research; whether the statement is open, settled, or controversial; supporting and opposing researchers and evidence; and references to the most relevant articles.
This saves Giuseppe days of work, and Isabelle gets a quick answer to her question. Furthermore, she can contribute to this global scientific knowledge base by publishing a nanopublication referring to her experiment that gave some weak evidence in favor the statement.
This takes her only a few minutes and might later have a positive impact on her citation record.

These examples show that informal representations of scientific statements (i.e. plain English sentences) are sufficient for many purposes. In fact, it would most certainly have been difficult for Giuseppe and Isabelle to come up with a formal representation for their hypothesis. It is known that even people who use ontology languages professionally often mix up such fundamental concepts as existential and universal quantification \cite{rector2004ekaw}. It is beyond question that formal representations have advantages that cannot be achieved with informal sentences, and we should express scientific claims in RDF form whenever possible. However, we think that in cases where formal representations are not practical (which could very well be the majority of cases), the scientific research community can derive substantial benefits from nanopublications with informal claims.

In the remainder of this paper, we will introduce a formalism for using English sentences in nanopublications. We discuss the constraints on such sentences and how they are generated, processed, and interlinked. We present evaluations on the ease of manually and automatically generating such nanopublications, and on sentence clustering for automatically linking related scientific claims.

\section{Background}
\label{sec:background}

There are only a few existing approaches of embedding scientific results as separate English sentences in formal structures. They are briefly outlined below, and contrasted with our approach in the next section.

SWAN (Semantic Web Applications in Neuromedicine) \cite{ciccarese2008bioinform} is an ontology that evolved from a web platform called \emph{Alzforum}, which has been used by the Alzheimer research community since 1996 to discuss their ideas and findings \cite{clark2007briefbioinf}. Similar to the approach to be presented here, SWAN provides a formal RDF-based scaffold for scholarly communication, while using informal English sentences to describe claims and hypotheses.
Another example is EXPO, an ontology for scientific experiments \cite{soldatova2006rsif}. In this model, each research hypothesis has both a formal definition and a natural language equivalent, and the latter typically has the form of a single English sentence.
As a third example, GeneRIF is a dataset describing gene and protein functions.\footnote{\url{http://www.ncbi.nlm.nih.gov/gene/about-generif}} Each GeneRIF entry consists of a gene ID, a publication reference, and --- most importantly --- a short English sentence of less than 255 characters describing a function of the given gene. We will use this dataset in one of our evaluations.

In addition to the nanopublication initiative, there are a number of related approaches of formally representing  scientific findings.
The Biological Expression Language (BEL) is ``a language for representing scientific findings in the life sciences in a computable form.''\footnote{\url{http://www.openbel.org}} It is embedded into a relatively complex scripting language called \emph{BEL Script}, where the formal statements can also be linked to sentences of the publication they were derived from.
Other approaches focus on hypotheses that are automatically generated \cite{soldatova2011biomedsem}.
A different application scenario is employed by an approach called \emph{structured digital abstracts} \cite{seringhaus2007bmcbioinf}, which should make formal representations of main scientific results sufficiently simple to require them directly from paper authors. These formal abstracts could be submitted, reviewed, and published together with their papers.

The particular kind of English sentences that our approach uses can be considered a controlled natural language (CNL) \cite{wyner2009cnlmain}. A CNL is a language that is based on a certain natural language, while being more restrictive concerning lexicon, grammar, and/or semantics. Previous work investigated the use of a formal CNL to write scientific abstracts that can be automatically translated into logic \cite{kuhn2006dils}. The CNL to be presented below is, however, of an essentially different type: It is much less restricted and is not designed for automatic interpretation.

\section{Approach}
\label{sec:approach}

\begin{figure}[tb]
\begin{center}
\includegraphics[scale=0.35]{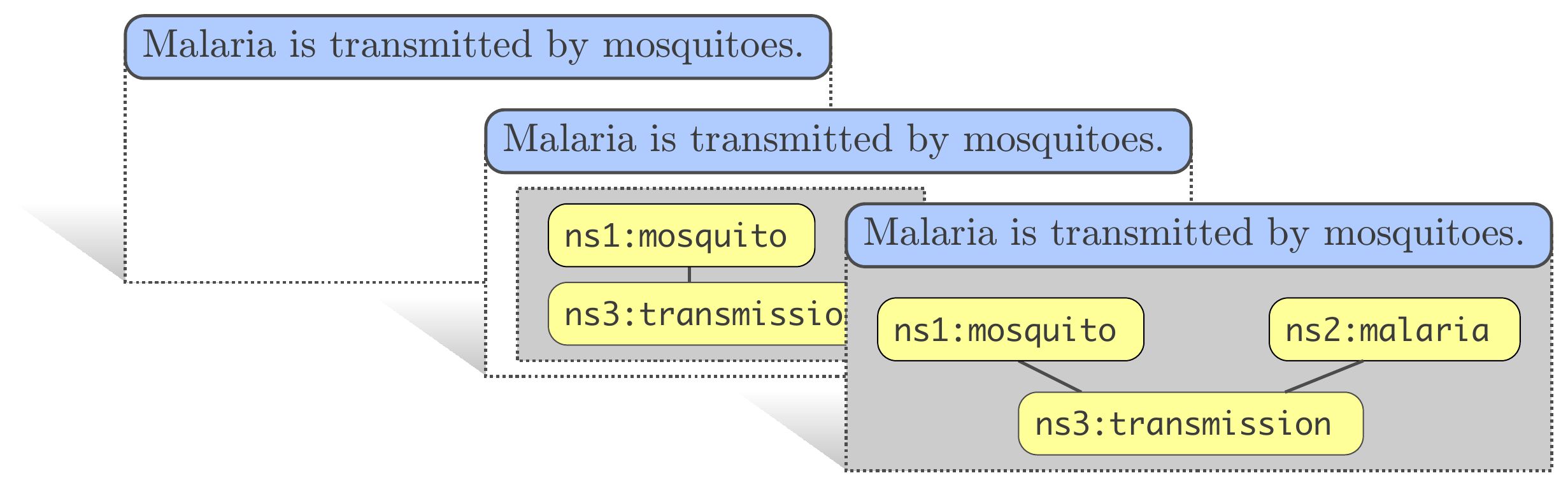}
\medskip
\hrule
\medskip
\includegraphics[scale=0.35]{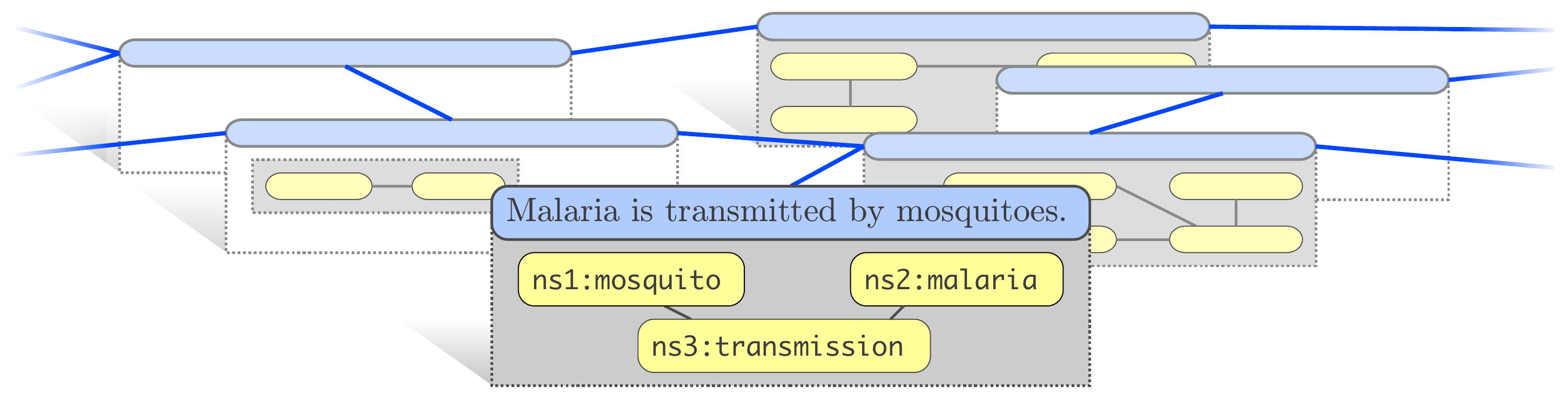}
\medskip
\hrule
\medskip
\includegraphics[scale=0.35]{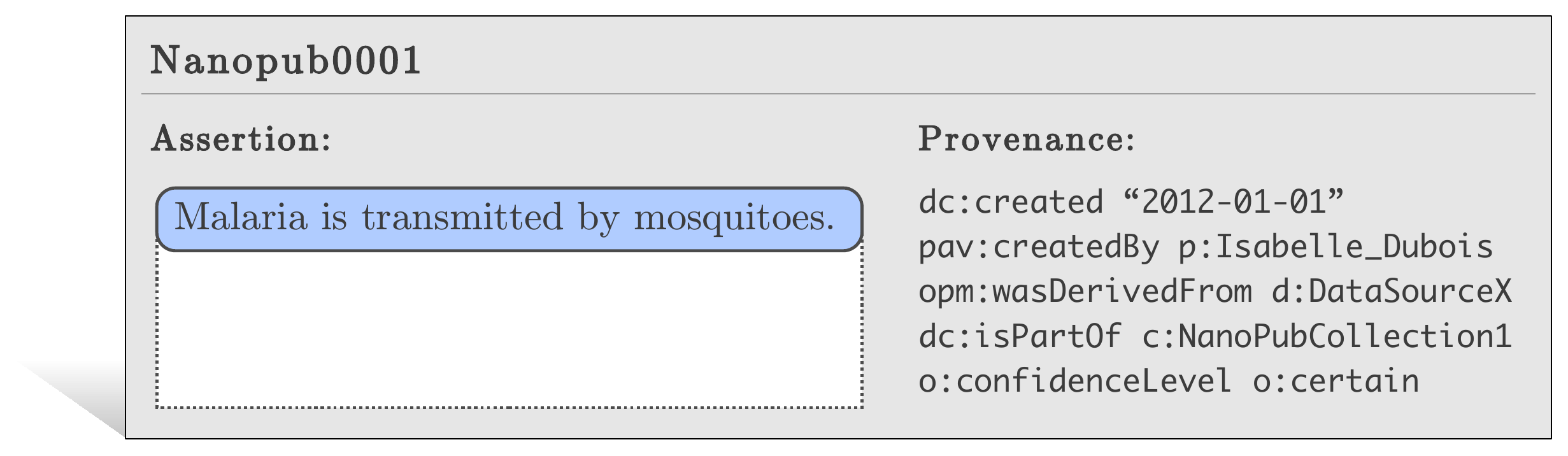}
\caption{In our approach, there is a continuum from formal to informal statements (top), which can be linked regardless of their level of formality (middle), and which can be asserted in nanopublications (bottom).}
\label{fig:linked}
\end{center}
\end{figure}

Our approach of using English sentences to describe scientific results differs from existing approaches --- such as SWAN, EXPO, and GeneRIF --- in the following four respects:
\begin{enumerate}
\item Our intended application range is very broad, covering science as a whole and beyond.
\item In our conceptualization, sentences exist independently from authors. In a certain sense, a sentence exists even if it has not yet been uttered by anybody. Conversely, a particular sentence might have been said by different persons at different points in time. None of these persons ``owns'' the sentence, but the sentence has an existence on its own and just happens to be mentioned (i.e. claimed, challenged, refuted, related, etc.) by people from time to time.
\item The sentences of our approach are not just any English sentences; we are more specific about what such sentences have to look like. We will introduce the concept of ``AIDA sentences,'' which can be considered a controlled natural language.
\item Our approach allows for a (quasi-)continuum from fully informal to fully formal statements. Natural sentences can be assigned partial or complete formal representations in the form of RDF graphs, combining the advantages of natural and formal representations.
\end{enumerate}
Number 2 might look like a purely philosophical issue, but it actually has very concrete consequences to our approach. For example, since sentences have their own independent existence, it is natural to give them URIs, making them first-class citizens in the RDF world. The formalization continuum of number 4 is shown in Figure \ref{fig:linked}, which also illustrates how statements can be interlinked regardless of their level of formality, and how they can be part of nanopublications. As a very simple (and in this sense untypical) example of a scientific claim, we borrow the sentence ``malaria is transmitted by mosquitoes'' from previous articles on nanopublications \cite{groth2010isu}.

\subsection{An Ocean of Nanopublications}

Before we move on to explain the details of our approach, let us describe the general nanopublication idea in some more detail. We deliberately present it from our own particular perspective, which embodies a view that is broader than the original one, but in a straightforward way. As motivated above, we believe that the application range of nanopublications could be much broader than what they were initially designed for (which is, by the way, very similar to the origins of the RDF standard). Basically, nanopublications could become the basis for the \emph{entire} Semantic Web. Whatever information one wants to share, it could be published in the form of one or more nanopublications. These can include scientific claims and experimental data, but also opinions, social relationships, events, properties of other nanopublications (``meta-nanopublications''), and much more.
In general, they are supposed to come from a number of channels, including the following:
\begin{enumerate}
\item Authors provide nanopublications for their own (scientific) results.
\item Users create meta-nanopublications by assessing, interlinking, and correcting existing nanopublications, claims, authors, and other relevant entities.
\item Curators generate nanopublications for results others have found.
\item Data mining (especially text mining) generates new nanopublications from existing unstructured data sources.
\item The data contained in existing structured data sources is exported into the nanopublication format.
\item Bots generate new nanopublications by crawling through the mass of existing ones and inferring obvious and not-so-obvious new relations.
\end{enumerate}
Channels 3, 4, and 5 are very important in the beginning to attain critical mass, but afterwards 1, 2, and 6 would gain importance. Channels 1, 2, 3, and 5 typically generate high-quality nanopublications, whereas 4 and 6 tend to have lower quality.
\begin{figure}[tb]
\begin{center}
\includegraphics[width=0.97\textwidth]{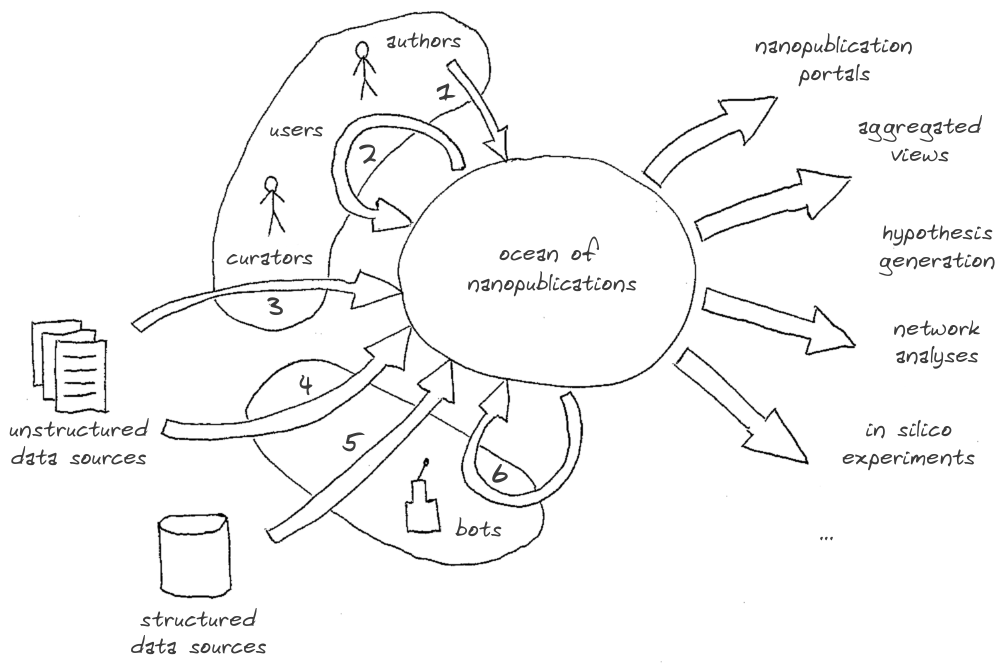}
\caption{Channels creating and using nanopublications}
\label{fig:flow}
\end{center}
\end{figure}
Figure \ref{fig:flow} shows these different channels and sketches some possible applications that consume nanopublications. In the middle of the picture, there is an ocean of nanopublications. At the moment, this is no more than a puddle, but the different channels should enlarge it to massive dimensions. A crucial question is whether these channels can produce enough nanopublications at the initial stage to let the ocean grow to a certain critical mass, at which point it would produce enough advantages for all participants to allow the system to run on its own. For that reason, the evaluations we will present below focus on the creation of nanopublications.

The agents that produce nanopublications can be humans or bots. We use the term \emph{bot} to denote ``robots without a body'' or ``named computer programs,'' i.e. agents that are made up only of software. Robot scientists \cite{king2009science} could become another important type of agent in the future.

\subsection{AIDA}

Let us turn now to the core of our approach, i.e. the particular kind of English sentences. They have to follow a scheme that we call AIDA (pronounced like the opera): the sentences have to be Atomic, Independent, Declarative and Absolute:
\begin{itemize}
\item \textbf{Atomic:} a sentence describing one thought that cannot be further broken down in a practical way
\item \textbf{Independent:} a sentence that can stand on its own, without external references like ``this effect'' or ``we''
\item \textbf{Declarative:} a complete sentence ending with a full stop that could in theory be either true or false
\item \textbf{Absolute:} a sentence describing the core of a claim ignoring the (un)cer\-tain\-ty about its truth and ignoring how it was discovered (no ``probably'' or ``evaluation showed that''); typically in present tense
\end{itemize}
The sentence ``malaria is transmitted by mosquitoes,'' which we have encountered above, is an example of an AIDA sentence.
The first three criteria basically reflect the nanopublication idea when applied to natural language instead of RDF.

The last AIDA criterion might look suspicious: After all, uncertainty is an essential aspect of scientific results. We are \emph{not} proposing to omit this aspect, but it should be recorded separately in the provenance part of the nanopublication and should not be part of the sentence. We do not have a concrete proposal at this point, but the ORCA model of uncertainty \cite{dewaard2012satbiswim} seems to be a very good candidate for integration. Once integrated, a user interface for creating AIDA nanopublications could look as follows:
\begin{quote}\small
{\Large\Square} We hypothesize that this statement might be true:\\
{\Large\XBox} We think this statement is probably true:\\
{\Large\Square} We think this statement is an established fact:
\smallskip\\
\fbox{\emph{Malaria is transmitted by mosquitoes. ~~~~~~~~~~~~~~~~~~~~~~~~~~~~~~~~~~}}
\end{quote}

As mentioned above, each AIDA sentence should get its own URI to make it a first-class citizen in the RDF world. String literals would not work out, as we want to establish relations between sentences, and RDF literals are not allowed in subject position of triples. An additional requirement is that the actual AIDA sentence should be extractable from its URI without consulting external resources, and vice versa. These requirements can be met by a straightforward URI encoding:
\begin{quote}
\small\url{http://purl.org/aida/Malaria+is+transmitted+by+mosquitoes.}
\end{quote}
No central authority is needed to approve new statements, but everybody can make up such URIs and immediately use them. We are aware that this goes against existing recommendations of keeping URLs opaque, but we think that when adding something essential such as a natural/formal continuum, a deliberate deviation from previous good practices is justified.

AIDA sentences can be interlinked by relations such as \texttt{\small has\-Same\-Mean\-ing}. The semantics of such relations is relatively straightforward for formal languages, but much less so for natural language, which is inherently vague and ambiguous. We employ a very pragmatic definition. In a nutshell, the semantics of an AIDA sentence is defined as the most frequent meaning English speakers assign to it.
More specifically, in order to find \emph{the} meaning of a given AIDA sentence, we mentally give it to all English speakers in the world, disregarding those who would say that they do not understand it. The remaining ones we ask (again mentally) about the most plausible meaning they would intuitively assign (without giving context information, as AIDA sentences are supposed to be independent). The most frequent of the resulting meanings is considered \emph{the} meaning of the AIDA sentence. On this basis, two sentences satisfy the \texttt{\small has\-Same\-Mean\-ing} relation, for example, if and only if we end up with equivalent meanings after going through the above mental exercise for each of them.

\subsection{Creating and Clustering AIDA Nanopublications}

The obvious channels that are supposed to provide us with AIDA nanopublications are channels 1, 3, and 4: Authors and curators manually write AIDA sentences, and text mining approaches automatically extract AIDA sentences from existing texts. However, bots can also produce AIDA nanopublications, inferring them from existing ones and interrelating them using NLP techniques (channel 5). In addition, users of nanopublication portals can link and correct existing AIDA sentences (channel 2). In the case of channel 6, we typically get formal representations ``for free,'' but having complementary AIDA sentences can still be helpful for humans to make sense of the respective claims.
The evaluation to be presented below focuses on channels 1, 3, 4, and 6.

As motivated in the introduction, the main benefit of AIDA nanopublications comes from interlinking them and relating them to other entities. The first problem we are facing is that a typical scientific statement can be expressed in more than one way. We, therefore, cannot expect that two AIDA sentences with the same meaning use exactly the same wording.
To solve this problem, we propose to use a mixture of automatic clustering and crowdsourcing. The clustering has the function of finding candidate sentences that seem to have similar or even identical meanings for a given AIDA sentence. Users of nanopublication portals can then filter out the false positives (ideally by single mouse clicks). These user responses would be published as nanopublications via channel 2. One of our studies to be introduced below shows results on the quality of automatic clustering of sentences.

\section{Implementation}
\label{sec:implementation}

Below, we introduce a prototype of a nanopublication portal and give some details on the RDF representations.

\subsection{Nanopublication Portal}

\begin{figure}[tb]
\begin{center}
\setlength{\fboxsep}{0mm}\fbox{\includegraphics[width=0.995\textwidth]{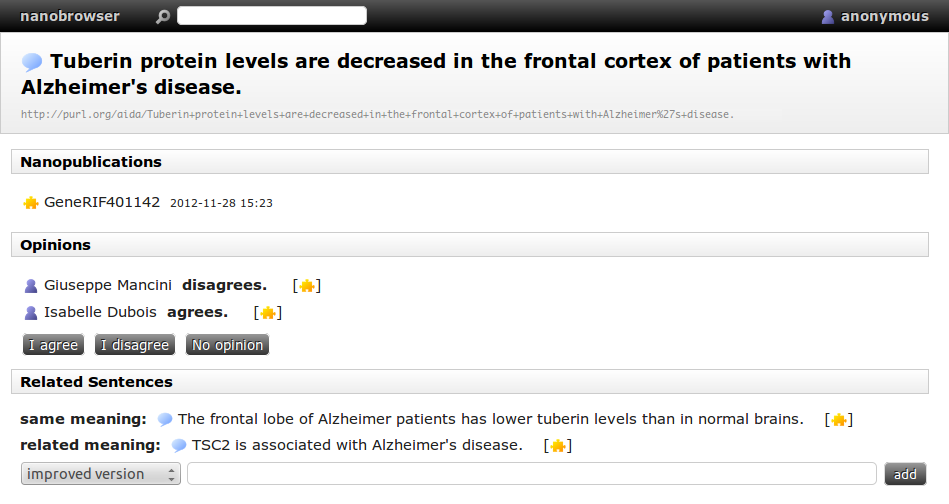}}
\caption{Screenshot of the nanobrowser interface}
\label{fig:nanobrowser}
\end{center}
\end{figure}

Figure \ref{fig:nanobrowser} shows a prototype of a nanopublication portal called \emph{nanobrowser} that we are developing to demonstrate our approach. It is based on Apache Wicket and the Virtuoso triple store, and its source code is available online.\footnote{\url{http://purl.org/nanobrowser}}
Nanobrowser is in fact more than just a browser and could be called a \emph{scientific\slash{}social\slash{}distributed\slash{}semantic wiki}. Users are presented small buttons such as ``I agree,'' which generate and publish meta-nanopublications by single mouse clicks. We still have to investigate what kind of opinions scientists would like to give. Something like ``I am (not) convinced'' might be better suited than ``I (dis)agree.'' In any case, users can in this way publish many nanopublications with little effort while browsing the knowledge base.

The screenshot of Figure \ref{fig:nanobrowser} shows a page that summarizes the available information about a particular scientific statement. The shown example was automatically extracted from GeneRIF as part of the evaluation to be described below. The page shows related sentences and opinions from researchers, each associated with the meta-nanopublication that established the respective relation (yellow jigsaw puzzle icon). Users can track down the origin of every piece of information to see how and by whom it was published.

\subsection{Extended Nanopublication Notation}

\newcommand{\rdfquotestyle}{\fontsize{8.5}{9.5}\selectfont}

Let us have a brief look under the hood, i.e. the actual RDF representation of nanopublications. The core part of a standard nanopublication is an \emph{assertion} in the form of a named graph \cite{carroll2005www}:
\begin{quote}
\rdfquotestyle
\begin{verbatim}
<> {
  :Pub1 np:hasAssertion :Pub1_Assertion .
  ...
}
:Pub1_Assertion { ... }
\end{verbatim}
\end{quote}
The curly brackets after \texttt{\small :Pub1\_Assertion} would contain the actual assertion in the form of a set of RDF triples. To allow for informal and underspecified assertions using AIDA sentences, we have to use a slightly more complex structure. With our approach, assertions consist of two subgraphs: a head and a body, where the body represents the actual (possibly unknown) formal representation:
\begin{quote}
\rdfquotestyle
\begin{verbatim}
<> {
  :Pub1 np:hasAssertion :Pub1_Assertion .
  :Pub1_Assertion np:containsGraph :Pub1_Assertion_Head .
  :Pub1_Assertion np:containsGraph :Pub1_Assertion_Body .
  ...
}
\end{verbatim}
\end{quote}
The head part is used to refer to different representations of the given assertion, such as the formal representation in the form of a named RDF graph or a natural representation in the form of an AIDA sentence:
\begin{quote}
\rdfquotestyle
\begin{verbatim}
:Pub1_Assertion_Head {
  :Pub1_Assertion
      npx:asSentence aida:Malaria+is+transmitted+by+mosquitoes. ;
      npx:asFormula :Pub1_Assertion_Body .
}
\end{verbatim}
\end{quote}
We can --- but we are not obliged to --- add a formalization of the given claim:
\begin{quote}
\rdfquotestyle
\begin{verbatim}
:Pub1_Assertion_Body { ... }
\end{verbatim}
\end{quote}
Partial representations can be defined in a straightforward way with the help of subgraphs, and we can use \texttt{\small rdf:about} to define that a certain entity must be part of a formalization without specifying a concrete triple:
\begin{quote}
\rdfquotestyle
\begin{verbatim}
:Pub1_Assertion_Body np:containsGraph :Pub1_Assertion_Body_Partial .
:Pub1_Assertion_Body rdf:about ns:malaria .
\end{verbatim}
\end{quote}
Later nanopublications can refer to \texttt{\small Pub1\_Assertion\_Body} to augment or correct the existing representation.

Overall, this extension is backwards compatible as long as \texttt{\small con\-tains\-Graph} relations are considered when retrieving the assertion triples, and allows for a uniform and general representation of informal, underspecified, and fully formal assertions.

\section{Evaluation}
\label{sec:evaluation}

It is obvious from Figure \ref{fig:flow} that there are many aspects to evaluate, most of which we have to leave to future work. As motivated above, we focus our evaluation on the left hand side of the picture, since this seems to be the critical part for the initial stage of our approach. The studies described below test some of the important aspects of the generation of AIDA nanopublications by both, humans and bots. Detailed supplementary material is available online.\footnote{\url{http://purl.org/tkuhn/aidapaper/supplementary}}

\subsection{Manual Generation of Nanopublications}

Our first evaluation tests aspects of channels 1 and 3, as described in Section \ref{sec:approach}: How easy or difficult is it for authors or curators to create nanopublications for their own or others' scientific results?

To that aim, we asked biomedical researchers to rewrite short texts from scientific abstracts as one or more AIDA sentences. In a sense, these participants resemble curators who are supposed to create nanopublications for existing scientific results. With respect to the lack of training and experience, however, they rather resemble authors who occasionally create nanopublications for their own results. Some of the tested aspects are therefore relevant to channel 1 and others to channel 3.

\subsubsection{Design.}

To get short original texts, we searched PubMed for articles with structured abstracts, i.e. abstracts that are divided into different parts like Introduction and Conclusions. According to our experience, the Conclusions section typically describes the general high-level scientific claims that correspond to the assertions of nanopublications. We took a random sample of PubMed abstracts that have a Conclusions section, excluding those that are not understandable without the broader context. Some of the resulting texts were shortened, so that each of them would lead to at most three AIDA sentences. We ended up with five such short texts.

We recruited 16 participants for this user study, all scientists with a background in biology and medicine. They had never heard of the AIDA concept before. They were directed to an online questionnaire that consisted of three parts: the first part briefly explained the AIDA concept; the second part showed the five short texts and asked for one to three AIDA sentences for each of them; the last part asked about the experienced difficulty of understanding the AIDA concept and of performing the rewriting tasks.

Below, one of the five short texts is shown as an example, together with two corresponding AIDA sentences as we got them from one of our participants:
\begin{quote}\small
\emph{Original text:} The results of this study showed that the hepatic reticuloendothelial function is impaired in cirrhotic patients, but the degree of impairment does not differ between patients with and without previous history of SBP. [PMID 11218245]
\medskip\par
\emph{AIDA 1:} The hepatic reticuloendothelial function is impaired in cirrhotic patients.
\smallskip\par
\emph{AIDA 2:} The degree of hepatic reticuloendothelial function impairment does not differ between cirrhotic patients with and without previous history of SBP.
\end{quote}

\subsubsection{Results.}

\begin{figure}[tp]
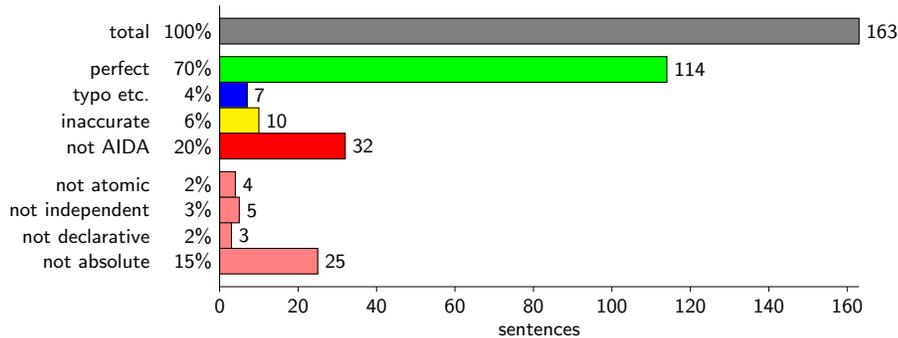

\begin{center}
\scalebox{0.85}{
\begin{bchart}[min=0,step=20,max=163,width=12.5cm,scale=0.8]
  \bcbar[label=total ~100\%,color=gray]{163}
  \smallskip
  \bcbar[label=perfect ~\phantom{1}70\%,color=green]{114}
  \bcbar[label=typo etc. ~\phantom{10}4\%,color=blue]{7}
  \bcbar[label=inaccurate ~\phantom{10}6\%,color=yellow]{10}
  \bcbar[label=not AIDA ~\phantom{1}20\%,color=red]{32}
  \smallskip
  \bcbar[label=not atomic ~\phantom{10}2\%,color=red!50]{4}
  \bcbar[label=not independent ~\phantom{10}3\%,color=red!50]{5}
  \bcbar[label=not declarative ~\phantom{10}2\%,color=red!50]{3}
  \bcbar[label=not absolute ~\phantom{1}15\%,color=red!50]{25}
  \bcxlabel{sentences}
\end{bchart}}
\vspace{-8mm}
\end{center}
\caption{Quality of the sentences created within the user study}
\label{fig:userstudyresults1}
\end{figure}

The 16 participants created 163 sentences in total. On average, they needed 15.3 minutes to complete the study. This means that an average sentence only required 90 seconds to be created, including the initial overhead to learn the AIDA concept.
We checked each of the 163 sentences whether it complied with the AIDA restrictions and whether it was an accurate representation of the original text. Some sentences contained minor mistakes, such as typos or missing copulas (e.g. ``X helpful in Y'' instead of ``X \emph{is} helpful in Y''). For the sentences that were not compliant with AIDA, we also checked which of the requirements they violated. The result for each sentence was based on two independent manual evaluations. Figure \ref{fig:userstudyresults1} shows the results.

70\% of the resulting sentences were perfect, which means that they complied with the AIDA restrictions and were accurate representations of the original texts. An additional 4\% were perfect, except that they contained typos and other minor mistakes. 6\% were AIDA-compliant but not accurate with respect to the original. 20\% violated one or several of the AIDA restrictions, mostly the absoluteness criterion (15\%).

As there is no straightforward way of assessing the recall for partially correct sentences, we calculated the ``recall for correct tasks'' by looking only at the tasks for which a particular participant wrote no incorrect sentence (except for typos): 96\% of these sentences covered all information of the initial text.


Next, we can have a look at the subjective experience of the participants, who had to specify their difficulty of understanding the AIDA concept and of performing the tasks. The possible answers were ``very difficult,'' ``difficult,'' ``easy,'' and ``very easy.'' All participants replied that understanding the concept of AIDA sentences was ``easy'' (but not ``very easy''). Nobody found it difficult or very difficult. The task of rewriting the short texts in AIDA format was of medium difficulty, with a tendency towards easy: ten out of the 16 participants found it ``easy''; the remaining six found it ``difficult.''


\subsection{Automatic Generation of Nanopublications}

Our second evaluation targets specific aspects of channels 4 and 6, as described in Section \ref{sec:approach}: How can we automatically extract nanopublications from text resources and then automatically relate them to each other?

For this part of the evaluation, we used the GeneRIF dataset,\footnote{\url{ftp://ftp.ncbi.nih.gov/gene/GeneRIF/generifs_basic.gz}} which contains sentences that describe the functions of genes and proteins. We evaluated the quality with which we can extract AIDA nanopublications from this dataset. Then, we investigated how well we can cluster them according to their similarity.

\subsubsection{Design.}

To automatically extract AIDA sentences, we tried to detect GeneRIF sentences that already follow the AIDA scheme. This was implemented as a set of simple regular expressions that filter out sentences that are unlikely to be AIDA-compliant. As many GeneRIF sentences start with phrases such as ``these results clearly indicated that'' or ``the authors propose that,'' and do therefore not adhere to the absoluteness criterion, we defined additional regular expressions to identify and remove such sentence beginnings, so that the remaining sentence texts could be treated as AIDA candidates. The resulting extraction program was a simple script containing these regular expressions.

During the development of this extraction program, the GeneRIF dataset as of September 2012 was used, which included roughly 750\,000 entries (including duplicate sentences). Upon completion of the extraction program in November 2012, we downloaded the latest version of GeneRIF, which had 16\,865 new entries. We then ran our extraction program on these new entries, which led to 4\,342 AIDA nanopublications. From these, we took a random sample of 250 unique sentences, which we manually checked for AIDA compliance.

As a next step, we extracted AIDA sentences from the entire GeneRIF dataset (119\,088 unique sentences) and added the ones we obtained from the user study described above (94 unique sentences). On average, each of the five user study tasks led to 18.8 unique statements, which were closely interrelated in terms of meaning but used different wording. We then applied a clustering algorithm on the combined set of sentences to evaluate the quality with which similar or equivalent sentences can be grouped, using the user study sentences as a kind of gold standard. As input for the clustering algorithm, we transformed the sentences into word vectors of \emph{tf-idf} values.

A plethora of unsupervised clustering methods have been developed in statistics and machine learning \cite{everitt1993cluster}. Most techniques require the user to define the number of clusters in advance, and those that do not, often require tuning of various parameters. Here, we use our own clustering method, specifically designed for sentence similarity, which deals with the large variation of neighbor density we observed in word vector space. Our algorithm goes as follows: (1) Given a point $X$ (i.e. a sentence) in our dataset, we build a model of its local environment $U_X$ by choosing a two-level set of nearest neighbors. (2) We repeatedly partition $U_X$ with $k$-means (with $k=3$) and consider the median distance $d_X$ of the elements of the cluster $C_X$ containing the base-point $X$. (3) If $d_X$ lies above a given threshold, point $X$ is considered an ``isolate'' and the cluster is discarded (this is to avoid ``loose'' clusters consisting of unrelated sentences, connected by low frequency words).

\subsubsection{Results.}

\begin{figure}[tp]
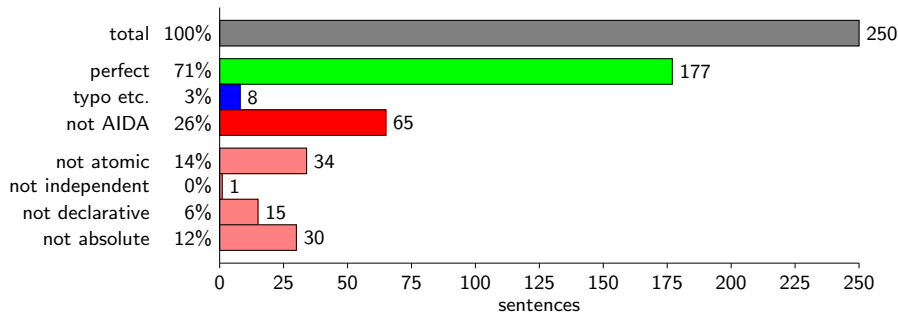

\begin{center}
\scalebox{0.85}{
\begin{bchart}[min=0,step=25,max=250,width=12.5cm,scale=0.8]
  \bcbar[label=total ~100\%,color=gray]{250}
  \smallskip
  \bcbar[label=perfect ~\phantom{1}71\%,color=green]{177}
  \bcbar[label=typo etc. ~\phantom{10}3\%,color=blue]{8}
  \bcbar[label=not AIDA ~\phantom{1}26\%,color=red]{65}
  \smallskip
  \bcbar[label=not atomic ~\phantom{1}14\%,color=red!50]{34}
  \bcbar[label=not independent ~\phantom{10}0\%,color=red!50]{1}
  \bcbar[label=not declarative ~\phantom{10}6\%,color=red!50]{15}
  \bcbar[label=not absolute ~\phantom{1}12\%,color=red!50]{30}
  \bcxlabel{sentences}
\end{bchart}}
\vspace{-8mm}
\end{center}
\caption{Quality of the sentences extracted from the GeneRIF dataset}
\label{fig:generifstudyresults1}
\end{figure}

Figure \ref{fig:generifstudyresults1} shows the results of our automatic extraction of AIDA nanopublications from the GeneRIF dataset. The general results look very similar to the ones from the user study. 71\% of the resulting AIDA sentences fully complied with the AIDA restrictions; an additional 3\% did so, but contained minor mistakes such as typos. In contrast with the user study, non-atomic sentences were relatively frequent (14\%). Creators of GeneRIF sentences are probably not encouraged to write a separate sentence for each claim, and non-atomicity is difficult to detect with simple regular expressions.

To evaluate the sentence clustering, we looked at the clusters for the sentences from the user study. We know that the set of sentences from a particular task has a large internal overlap in terms of meaning (which does not mean that they are all similar, as the respective text may describe more than one claim). Our results show that sentences from the same task indeed end up in the same clusters. On average, 99.2\% of the other objects that such a sentence encountered in its cluster were sentences from the same task. Furthermore, 84\% of the sentences were connected to at least one other sentence from the same task. Below is an example of two sentences that our clustering algorithm successfully connected. They convey the same meaning, but are quite different on the surface level:
\begin{itemize}\small
\item Hepatic reticuloendothelial function is impaired to the same degree in cirrhotic patients with or without a previous history of SBP.
\item History of spontaneous bacterial peritonitis does not affect impairment of hepatic reticuloendothelial function in cirrhotic patients.
\end{itemize}

For both evaluations on automatic processing (i.e. extraction and clustering), we applied very simple methods and we assume that we can achieve even better results with state-of-the-art NLP techniques, language resources, and ontologies.




\section{Conclusions}
\label{sec:conclusions}

The pace of modern science is such that it is very difficult to keep track of the latest research results. Our approach addresses this problem by allowing researchers to easily access and communicate research hypotheses, claims, and opinions within the existing nanopublication framework. Representing scientific claims as AIDA sentences makes the nanopublication concept much more flexible and significantly widens its practical applicability.
Our results show that scientists are able to efficiently produce high-quality AIDA nanopublications, that it is feasible to extract such nanopublications from existing text resources, and that it is possible to cluster them by sentence similarity. Together, these findings suggest that our approach is practical, and that it may assist the nanopublication initiative to attain critical mass.

\bibliography{nanopub}
\bibliographystyle{plain}

\end{document}